\documentclass[
    final            % use final for the camera ready runs
  4apaper]
  {aipproc}

\layoutstyle{8x11single}

    \usepackage{bm}

\newcommand\beq{\begin{equation}}
\newcommand\eeq{\end{equation}}
\newcommand\beqa{\begin{eqnarray}}
\newcommand\eeqa{\end{eqnarray}}
\newcommand{\nn}{\nonumber\\}

\newcommand{\text}{\mathrm}

\newcommand{\rr}{\mathbf{r}}

\newcommand{\VV}{\mathbf{V}}

\newcommand{\el}{\text{el}}

\newcommand{\inel}{\text{inel}}

\newcommand{\ro}{\text{rough}}

\newcommand{\cc}{\mathbf{v}}

\newcommand{\ww}{\bm{\omega}}

\newcommand{\esn}{\alpha}

\newcommand{\est}{\beta}

\newcommand{\q}{\kappa}

\newcommand{\ft}{f^\text{tr}}
\newcommand{\fr}{f^\text{rot}}

\newcommand{\Tt}{T^\text{tr}}

\newcommand{\Tr}{T^\text{rot}}
\newcommand{\Trb}{\overline{T}^\text{rot}}

\newcommand{\zt}{\xi^\text{tr}}

\newcommand{\zr}{\xi^\text{rot}}

\newcommand{\tr}{\text{tr}}
\newcommand{\rot}{\text{rot}}

\begin{document}

\title{A Bhatnagar--Gross--Krook-like Model Kinetic Equation for a Granular Gas of Inelastic Rough Hard Spheres}


\author{Andr\'es Santos}{
  address={Departamento de F\'{\i}sica, Universidad de
Extremadura, E-06071 Badajoz, Spain} }

\begin{abstract}
The Boltzmann collision operator for a dilute granular gas of inelastic rough hard spheres is much more intricate than its counterpart for inelastic smooth spheres. Now the one-body distribution function depends not only on the translational velocity $\mathbf{v}$ of the center of mass  but also on the angular velocity $\bm{\omega}$ of the particle. Moreover, the collision rules couple $\mathbf{v}$ and  $\bm{\omega}$,  involving not only the coefficient of normal restitution $\alpha$ but also the coefficient of tangential restitution $\beta$.
The aim of this paper is to propose an extension to  inelastic rough particles of a Bhatnagar--Gross--Krook-like kinetic model previously proposed for inelastic smooth particles. The Boltzmann collision operator is replaced by the sum of three terms representing: (i) the relaxation to a two-temperature local equilibrium distribution, (ii)  the action of a  nonconservative drag force $\mathbf{F}$ proportional to $\mathbf{v}-\mathbf{u}$ ($\mathbf{u}$ being the flow velocity), and (iii) the action of a nonconservative torque $\mathbf{M}$ equal to a linear combination of $\bm{\omega}$ and $\bm{\Omega}$ ($\bm{\Omega}$ being  the mean angular velocity). The three coefficients in $\mathbf{F}$ and $\mathbf{M}$ are fixed to reproduce the Boltzmann collisional rates of change of  $\bm{\Omega}$ and of the two granular temperatures (translational and rotational). A simpler version of the model is also constructed in the form of  two coupled kinetic equations for the translational and rotational velocity distributions. The kinetic  model is applied to the simple shear flow steady state and the combined influence of $\alpha$ and $\beta$ on the shear and normal stresses and on the translational velocity distribution function is analyzed.
\end{abstract}

\maketitle

%%%%%%%%%%%%%%%%%%%%%%%%%%%%%%%%%%%%%%%%%%%%
%% MAINMATTER
%%%%%%%%%%%%%%%%%%%%%%%%%%%%%%%%%%%%%%%%%%%%

\section{Introduction and motivation}
As is well known, the master equation for the dynamics of an ordinary gas in the rarefied regime is the Boltzmann equation \cite{C88}. In the particular case of hard spheres, it reads
\beq
\partial_t f(\rr,\cc;t)+\cc\cdot \nabla f(\rr,\cc;t)=J^\el[\cc|f(\rr,\cdot;t),f(\rr,\cdot;t)],
\label{1.1}
\eeq
where $f(\rr,\cc;t)$ is the one-particle velocity distribution function and
\beq
J^\el\left[{\bf v}_{1}|f(\rr,\cdot;t),f(\rr,\cdot;t)\right]=\sigma^{2} \int
d{\bf v}_{2}\int d\widehat{\bm{\sigma}} \, \Theta\left(\mathbf{g}\cdot\widehat{\bm{\sigma}}\right)\left(\mathbf{g}\cdot\widehat{\bm{\sigma}}\right)\left[
f(\rr,{\bf v}_{1}'';t)f(\rr,{\bf v}_{2}'';t)-f(\rr,{\bf v}_{1};t)f(\rr,{\bf
v}_{2};t)\right]
\label{1.2}
\eeq
is the Boltzmann collision operator. Here, $\sigma$ is the diameter of a sphere, $\Theta(x)$ is Heaviside's step function, $\widehat{\boldsymbol{\sigma}}$ is a unit vector
directed along the centers of the two colliding particles, $\mathbf{g}=\mathbf{v}_1-\mathbf{v_2}$ is the relative velocity, and  the double primes on the velocities denote the
initial values $\{{\bf v}_{1}'', {\bf v}_{2}''\}$ that
lead to $\{{\bf v}_{1},{\bf v}_{2}\}$ following an \emph{elastic}  binary collision:
\beq
\label{1.3}
{\bf v}_{1}''={\bf v}_{1}-(\widehat{\boldsymbol{\sigma}}\cdot {\bf
g})\widehat{\boldsymbol {\sigma}}, \quad
{\bf v}_{2}''={\bf
v}_{2}+
(\widehat{\boldsymbol{\sigma}}\cdot {\bf
g})\widehat{\boldsymbol{\sigma}}.
\eeq
The basic properties of the Boltzmann collision operator are
\beq
\int d\mathbf{v}\, J^\el[\mathbf{v}|f,f]=0,\quad \int d\mathbf{v}\, \mathbf{v}J^\el[\mathbf{v}|f,f]=\mathbf{0},
\label{1.4}
\eeq
\beq
\int d\mathbf{v}\, \cc^2 J^\el[\mathbf{v}|f,f]=0,
\label{1.5}
\eeq
\beq
J^\el[\mathbf{v}|f_0,f_0]=0\Leftrightarrow f_0(\cc)= n\left(\frac{m}{2\pi T}\right)^{3/2}\exp\left(-\frac{m V^2}{2T}\right) ,
%e^{-m\left(\mathbf{v}-\mathbf{u}\right)^2/2T}.
\label{1.6}
\eeq
where $\VV\equiv \cc-\mathbf{u}$ is the peculiar velocity. Henceforth, for the sake of notation simplicity, the arguments $\rr$ and $t$ will be implicitly understood and generally omitted.
Equations \eqref{1.4} and \eqref{1.5} express the collisional conservation of  mass, momentum, and energy, while Eq.\ \eqref{1.6} indicates that collisions do not change the local equilibrium distribution function. In the latter quantity, $m$ is the mass of a particle and  $n=\int d\cc \, f(\cc)$, $\mathbf{u}=n^{-1}\int d\cc \,\cc f(\cc)$, and $T=({m}/{3n})\int d\cc \, V^2f(\cc)$ are the local number density, flow velocity, and temperature, respectively. Note that, as usually done  in the literature on granular gases, the Boltzmann constant has been absorbed in the definition of temperature, so that the latter quantity has dimensions of energy.

The mathematical intricacy of the Boltzmann operator \eqref{1.2} makes it difficult to get explicit nonequilibrium solutions of Eq.\ \eqref{1.1}, especially beyond the scope of the Chapman--Enskog method. This has motivated the proposal of simplified model kinetic equations \cite{C88} where $J^\el[\cc|f,f]$ is replaced by a more tractable term $K^\el[\cc|f]$ that preserves Eqs.\ \eqref{1.4}--\eqref{1.6}.
The simplest and most celebrated of these model kinetic equations is perhaps the one proposed in 1954 by Bhatnagar, Gross, and Krook (BGK) and, independently, by Welander \cite{BGK54}. It consists of the replacement
\beq
J^\el[\cc|f,f]\to K^\el[\cc|f]=-\nu\left[f(\cc)-f_0(\cc)\right], \quad \nu=\frac{16}{5}\sigma^2 n\sqrt{\pi T/m},
\label{1.8}
\eeq
where $\nu(\rr,t)$ is an effective velocity-independent collision frequency. Its expression is not fixed by Eqs.\ \eqref{1.4}--\eqref{1.6} and thus it can be freely chosen to  optimize further agreement with the original Boltzmann equation. The explicit expression in Eq.\ \eqref{1.8}  guarantees that the Navier--Stokes shear viscosity obtained from the BGK model coincides with that obtained from the Boltzmann equation in the first Sonine approximation \cite{C88}.
The BGK kinetic model has proven to be much more reliable than its formulation might anticipate, even in far from equilibrium states \cite{GS03}.

Needless to say, the complexity of the Boltzmann equation increases when it describes a granular gas made of \emph{inelastic} hard spheres \cite{BP04}. In that case, the Boltzmann collision operator is given by
\beq
J^\inel\left[{\bf v}_{1}|f,f\right]=\sigma^{2} \int
d{\bf v}_{2}\int d\widehat{\bm{\sigma}} \, \Theta\left(\mathbf{g}\cdot\widehat{\bm{\sigma}}\right)\left(\mathbf{g}\cdot\widehat{\bm{\sigma}}\right)\left[
\alpha^{-2}f({\bf v}_{1}'')f({\bf v}_{2}'')-f({\bf v}_{1})f({\bf
v}_{2})\right],
\label{1.10}
\eeq
where $\alpha<1$ is the coefficient of normal restitution (here assumed to be constant) and Eq.\ \eqref{1.3} must be replaced by
\beq
\label{1.11}
{\bf v}_{1}''={\bf v}_{1}-\frac{1+\alpha
^{-1}}{2}(\widehat{\boldsymbol{\sigma}}\cdot {\bf
g})\widehat{\boldsymbol {\sigma}}, \quad
{\bf v}_{2}''={\bf
v}_{2}+\frac{1+\alpha
^{-1}}{2}
(\widehat{\boldsymbol{\sigma}}\cdot {\bf
g})\widehat{\boldsymbol{\sigma}}.
\eeq
Mass and momentum are still conserved, so that Eq.\ \eqref{1.4} is also satisfied by $J^\inel$. On the other hand, the kinetic energy is dissipated by collisions, so that one has
\beq
\zeta\equiv -\frac{m}{3nT}\int d\cc\, \cc^2 J^\inel[\cc|f,f]>0
\label{1.12}
\eeq
instead of Eq.\ \eqref{1.5}. Equation \eqref{1.12} defines the so-called \emph{cooling rate} $\zeta$ as a functional of $f$. Moreover, inelasticity prevents the system from reaching an equilibrium state and so Eq.\ \eqref{1.6} is not fulfilled by $J^\inel$. Instead, an undriven homogeneous system freely cools down according to $\partial_t T=-\zeta_h T$, where $\zeta_h$ is the cooling rate corresponding to the so-called homogeneous cooling state $f_h$, which is a \emph{similarity} solution of the Boltzmann equation depending on time only through the temperature. As a consequence, the role of $f_0$ is played by $f_h$ and Eq.\ \eqref{1.6} is replaced by \cite{BDS99}
\beq
J^\inel[\cc|f_h,f_h]=\frac{\zeta_h}{2}\frac{\partial}{\partial \cc}\cdot\left(\cc f_h\right).
\label{1.13}
\eeq

As in the elastic case, it is desirable to construct a kinetic model for granular gases by replacing the true operator $J^\inel[\cc|f,f]$ with a simpler term $K^\inel[\cc|f]$. Inspired by the BGK model \eqref{1.8} and by the properties \eqref{1.12} and \eqref{1.13}, the following kinetic model was proposed in Ref.\ \cite{BDS99}:
\beq
J^\inel[\cc|f,f]\to K^\inel[\cc|f]=-\lambda(\alpha)\nu\left[f(\cc)-f_h(\cc)\right]+\frac{\zeta}{2}\frac{\partial}{\partial \cc}\cdot\left[\VV f(\cc)\right],
\label{1.14}
\eeq
where $f_h$ is here the  \emph{local} form of the homogeneous cooling state solution, obtained from the latter by replacing $\cc\to \VV$ and by replacing the temperature and density with their local values for the nonequilibrium state under consideration. As also happened in the original BGK model \eqref{1.8}, the modified collision frequency $\lambda\nu$ remains a free parameter. By keeping  $\nu$ as given by Eq.\ \eqref{1.8}, the factor $\lambda(\alpha)$ can be chosen, for instance, to optimize agreement with the viscosity, the thermal conductivity, or the self-diffusion coefficient for the Boltzmann equation \cite{BDS99,SA05}. The latter criterion simply yields $\lambda(\alpha)=(1+\alpha)/2$. The kinetic model \eqref{1.14} is still too complicated because it requires, on the one hand, the solution of Eq.\ \eqref{1.13} and, on the other hand, the cooling rate as a functional of $f$ through Eq.\ \eqref{1.12}. In the spirit of a simple kinetic model both conditions can be relaxed \cite{BDS99} by substituting in Eq.\ \eqref{1.14} the unknown local distribution function $f_h$ with the local equilibrium distribution $f_0$ and the detailed cooling rate $\zeta$ with its local equilibrium estimate
$\zeta\to \frac{5}{12}(1-\alpha^2)\nu$.

While describing a granular gas, both the Boltzmann operator \eqref{1.10} and its associated BGK-like model \eqref{1.14} assume that the inelastic hard spheres are \emph{smooth}. However, there exist a number of relevant effects in granular gases where roughness plays a crucial role \cite{MSS04,Z06,KBPZ09}.
The main purpose of this paper is to extend the BGK-like model \eqref{1.14} to the case of inelastic \emph{rough} hard spheres characterized, apart from $\alpha$, by a coefficient of tangential restitution $\beta$. The situation is now much more complex since, in addition to the translational degrees of freedom, there also exist rotational ones \cite{Z06,KBPZ09}. In fact, one can define two independent granular temperatures, $\Tt$  and $\Tr$, and thus there are two associated energy production rates, $\zt$ and $\zr$, instead of a single cooling rate $\zeta$.
The proposed kinetic model  will be further simplified by focusing on the translational distribution function. This latter model will be applied and solved to the steady simple or uniform shear flow.

\section{The Boltzmann equation for inelastic rough hard spheres\label{sec2}}
Let us consider a granular gas made of inelastic rough hard spheres characterized by a constant coefficient of normal restitution $\alpha$ and a constant coefficient of tangential restitution $\beta$. The coefficient $\alpha$ ranges from $\alpha=0$ (collisions perfectly inelastic) to $\alpha=1$ (collisions perfectly elastic), while the coefficient $\beta$ ranges from $\beta=-1$ (spheres perfectly smooth) to $\beta=1$ (spheres perfectly rough). Except if $\alpha=1$ and $\beta=\pm 1$, energy is dissipated upon collisions. The one-particle distribution function $f(\rr,\cc, \ww;t)$ depends not only on the translational velocity $\cc$ of the center of mass but also on the angular velocity $\ww$. In the dilute regime $f$ obeys the Boltzmann equation
\beq
\partial_t f(\rr,\cc,\ww;t)+\cc\cdot \nabla f(\rr,\cc,\ww;t)=J^\ro[\cc,\ww|f(\rr,\cdot,\cdot;t),f(\rr,\cdot,\cdot;t)],
\label{2.1}
\eeq
where the collision operator is
\beq
J^\ro\left[{\bf v}_{1},\ww_1|f,f\right]=\sigma^{2} \int
d{\bf v}_{2}\int d\ww_2\int d\widehat{\bm{\sigma}} \, \Theta\left(\mathbf{g}\cdot\widehat{\bm{\sigma}}\right)\left(\mathbf{g}\cdot\widehat{\bm{\sigma}}\right)
\left[
(\alpha\beta)^{-2}f({\bf v}_{1}'',\ww_1'')f({\bf v}_{2}'',\ww_2'')-f({\bf v}_{1},\ww_1)f({\bf
v}_{2},\ww_2)\right].
\label{2.2}
\eeq
The restituting collision rules are \cite{Z06,SKG10}
\beq
\label{2.3}
{\bf v}_{1}''={\bf v}_{1}-\mathbf{C}, \quad
{\bf v}_{2}''={\bf
v}_{2}+\mathbf{C},\quad \ww_1''=\ww_1-\frac{2}{\sigma \kappa}\widehat{\bm{\sigma}}\times \mathbf{C},\quad \ww_2''=\ww_2-\frac{2}{\sigma \kappa}\widehat{\bm{\sigma}}\times \mathbf{C},
\eeq
where
\beq
\kappa\equiv \frac{4I}{m\sigma^2},\quad \mathbf{C}\equiv
\frac{1+\alpha
^{-1}}{2}(\widehat{\boldsymbol{\sigma}}\cdot {\bf
g})\widehat{\boldsymbol {\sigma}}+\frac{\kappa}{1+\kappa}\frac{1+\beta^{-1}}{2}\left[\mathbf{g}-(\widehat{\boldsymbol{\sigma}}\cdot {\bf
g})\widehat{\boldsymbol {\sigma}}-\frac{\sigma}{2}\widehat{\bm{\sigma}}\times\left(\ww_1+\ww_2\right)\right],
\label{2.4}
\eeq
$I$ being the moment of inertia. The value of $\q$ depends on the mass distribution within the sphere and runs from the extreme values $\q=0$ (mass concentrated on the center) to $\q=\frac{2}{3}$ (mass concentrated on the surface); if the mass is uniformly distributed, then $\q=\frac{2}{5}$. Mass and linear momentum are conserved by collisions, so one has
\beq
\int d\mathbf{v}\int d\ww\, J^\ro[\mathbf{v},\ww|f,f]=0,\quad \int d\mathbf{v}\int d\ww\, \mathbf{v}J^\ro[\mathbf{v},\ww|f,f]=\mathbf{0}.
\label{2.5}
\eeq
On the other hand, except in the perfectly smooth case ($\beta=-1$), collisions tend to decrease the mean angular velocity $\bm{\Omega}(\rr,t)$. This can be characterized by a ``de-spinning'' rate $\zeta_\Omega$ defined by
\beq
\int d\cc\int d\ww\, \ww J^\ro[\cc,\ww|f,f]=-\zeta_\Omega n\bm{\Omega},\quad \bm{\Omega}=\frac{1}{n}\int d\cc\int d\ww\, \ww f(\cc,\ww).
\label{2.9}
\eeq
In general, neither the translational nor the rotational kinetic energies are conserved by collisions. This can be characterized by the partial \emph{energy production rates}
\beq
\zt\equiv -\frac{m}{3n\Tt}\int d\cc \int d\ww\,\cc^2 J^\ro[\cc,\ww|f,f],\quad \zr\equiv -\frac{I}{3n\Tr}\int d\cc \int d\ww\,\omega^2 J^\ro[\cc,\ww|f,f].
\label{2.6}
\eeq
Here, $\Tt=({m}/{3n})\int d\cc \int d\ww\, V^2f(\cc,\ww)$ and $\Tr=({I}/{3n})\int d\cc \int d\ww\, \omega^2f(\cc,\ww)$ are the translational and rotational temperatures, respectively.
In general, collisions produce a transfer of energy between the translational and rotational degrees of freedom and, consequently, the energy production rates $\zt$ and $\zr$ do not have a definite sign. On the other hand, unless $\alpha=1$ and $\beta=\pm 1$, the total energy (translational plus rotational) is dissipated and thus the net \emph{cooling rate}
\beq
\zeta=\frac{1}{2 T}\left(\Tt \zt+\Tr \zr\right), \quad T=\frac{\Tr+\Tt}{2},
\label{2.8}
\eeq
is positive definite, where $T$ is the total temperature.

Note that, instead of $\Tr$, one could have alternatively adopted
$
 \Trb=({I}/{3n})\int d\cc \int d\ww\, \left(\bm{\omega}-\bm{\Omega}\right)^2f(\cc,\ww)=\Tr\left(1-X\right)$, with
$ X\equiv {\q m\sigma^2\Omega^2}/{12\Tr}$,
as the definition of the rotational temperature. Its associated production rate is
$
\overline{\xi}^\rot=({\zr-2\zeta_\Omega X})/({1-X})$.
However, the disadvantage of this alternative choice  is that, in contrast to $\zeta$,  the ``cooling'' rate  $\overline{\zeta}=\left(\zeta T-2\zeta_\Omega  X\right)/\overline{T}$ associated with the alternative total temperature  $\overline{T}=\frac{1}{2}(\Tt+\Trb)=T-\frac{1}{2}\Tr X$ is not positive definite and in fact becomes negative in the perfectly elastic and rough case ($\alpha=1$, $\beta=1$).

Equations \eqref{2.9}--\eqref{2.8} define the de-spinning rate ($\zeta_\Omega$), the energy production rates ($\zt$, $\zr$), and the cooling rate ($\zeta$)  as functionals of the velocity distribution function $f$. They can be estimated in terms of the local values of $n$, $\Tt$, $\Tr$, and $\bm{\Omega}$ by the replacements
\beq
f(\cc,\ww)\to \ft(\cc)\fr(\ww),\quad \ft(\cc)\to \ft_0(\cc),
\label{2.11}
\eeq
where
\beq
\ft(\cc)=\int d\ww\, f(\cc,\ww),\quad \fr(\ww)=\frac{1}{n}\int d\cc\, f(\cc,\ww)
\label{2.10}
\eeq
are \emph{marginal} distribution functions and $\ft_0$ is given by Eq.\ \eqref{1.6} with $T\to \Tt$. The results are \cite{Z06,SKG10}
\beq
\zeta_\Omega=\frac{5}{6}\frac{1+\beta}{1+\kappa}\nu,\quad \nu\equiv\frac{16}{5}\sigma^2 n\sqrt{\pi \Tt/m}.
\label{2.15}
\eeq
\beq
\zt=\frac{5}{12}\left[1-\esn^2+\frac{\q}{1+\q}\left(1-\est^2\right)+\frac{\q}{(1+\q)^2}\left(1+\est\right)^2\left(1-\frac{\Tr(1+X)}{\Tt}\right)\right]\nu,
\label{2.12}
\eeq
\beq
\zr=\frac{5}{12}\frac{1+\est}{1+\q}\frac{\Tt}{\Tr}\left[(1-\est)\frac{\Tr(1+X)}{\Tt}-\frac{\q}{1+\q}\left(1+\est\right)\left(1-\frac{\Tr(1+X)}{\Tt}\right)\right]\nu,
\label{2.13}
\eeq
\beq
\zeta=\frac{5}{12}\frac{\Tt}{\Tt+\Tr}\left[1-\esn^2+\frac{1-\est^2}{1+\q}\left(\q+\frac{\Tr(1+X)}{\Tt}\right)\right]\nu.
\label{2.14}
\eeq

Before closing this section it is worthwhile noting that the vanishing of the second integral in \eqref{2.5} is not verified in the case of the operator $J^\ro_{12}[f_1,f_2]$ describing the collisions of tagged particles (label 1) with untagged particles (label 2). Under conditions milder  than Eq.\ \eqref{2.11} one has \cite{SKG10}
\beq
\int d\cc_1 \int d\ww_1 \,\cc_1 J^\ro_{12}[\cc_1,\ww_1|f_1,f_2] \approx
\lambda(\alpha,\beta)
\int d\cc_1 \int d\ww_1 \,\cc_1 J^\el_{12}[\cc_1,\ww_1|f_1,f_2], \quad
\lambda(\alpha,\beta)
\equiv\frac{1+\alpha}{2}+\frac{\q}{1+\q}\frac{1+\beta}{2}.
\label{2.16}
\eeq
According to Eq.\ \eqref{2.16}, the average collisional transfer of momentum from component 2 to component 1 in the inelastic rough case is $\lambda(\alpha,\beta)$ times the value in the elastic smooth case. This property is relevant for self-diffusion problems.

\section{Kinetic modeling\label{sec3}}

\subsection{Joint distribution function}
As in the smooth-sphere case, both elastic [cf.\ Eq.\ \eqref{1.8}] and inelastic [cf.\ Eq.\ \eqref{1.14}], the idea behind a kinetic model is the replacement of the complex Boltzmann collision operator $J^\ro[f,f,]$ by a much simpler term $K^\ro[f,f]$ that otherwise retains a number of basic physical conditions. As a natural extension of Eq.\ \eqref{1.14}, the BGK-like model proposed here for the joint distribution $f$ is
\beqa
J^\ro[\cc,\ww|f,f]\to K^\ro[\cc,\ww|f]&=&-\lambda(\alpha,\beta)\nu\left[f(\cc,\ww)-f_0^\tr(\cc)f_0^\rot(\ww)\right]+\frac{\zt}{2}\frac{\partial}{\partial \cc}\cdot\left[\VV f(\cc,\ww)\right]\nn
&&
+\frac{1}{2}\frac{\partial}{\partial \ww}\cdot\left\{\left[2\zeta_\Omega\bm{\Omega} +{\overline{\xi}^\rot}\left(\ww-\bm{\Omega}\right)\right]f(\cc,\ww)\right\},
\label{2.17}
\eeqa
where
\beq
f_0^\tr(\cc)\equiv n\left(\frac{m}{2\pi \Tt}\right)^{3/2}\exp\left(-\frac{mV^2}{2\Tt}\right)
,\quad
f_0^\rot(\ww)\equiv \left(\frac{I}{2\pi\Trb}\right)^{3/2}\exp\left[-
\frac{I\left(\ww-\bm{\Omega}\right)^2}{2\Trb}\right]
\label{2.18}
\eeq
are the local equilibrium distributions at independent temperatures.
The second term on the right-hand side of Eq.\ \eqref{2.17} can be interpreted as representing the action of an external drag force $\mathbf{F}=-\frac{m}{2}{\zt}\VV$. Likewise, the third term represents an external torque of the form $\mathbf{M}=-\frac{I}{2}\left[2\zeta_\Omega\bm{\Omega} +\overline{\xi}^\rot\left(\ww-\bm{\Omega}\right)\right]$.
By construction, the model \eqref{2.17} complies with the exact properties \eqref{2.5}--\eqref{2.6}, regardless of the choice of the effective collision frequency $\lambda \nu$. Based on the property \eqref{2.16}, here the adopted choice for  $\nu$ and $\lambda$ is given by Eqs.\ \eqref{2.15} and
\eqref{2.16}, respectively. As for $\zeta_\Omega$, $\zt$, and $\zr$, they are explicitly given by Eqs.\ \eqref{2.15}--\eqref{2.13}.

\subsection{Marginal distributions}
The kinetic model \eqref{2.17} represents a significant simplification with respect to the original Boltzmann collision operator \eqref{2.2}. However, it is still too complicated a model if one wants to apply it to situations where the BGK-like model \eqref{1.14} for inelastic smooth spheres allows for exact solutions \cite{SA05,BRM97,TTMGSD01}. In order to get explicit results that might be useful to assess the influence of roughness on the basic physical properties, it seems desirable to use Eq.\ \eqref{2.17} as the starting point for an even simpler kinetic model.

More explicitly,  we consider now the kinetic equations for the marginal distributions defined by Eq.\ \eqref{2.10}. When the replacement
\eqref{2.17} is inserted into Eq.\ \eqref{2.1}  and integration over $\ww$ or over $\cc$ are carried out, one obtains
\beq
\partial_t f^\tr(\cc)+\cc\cdot \nabla f^\tr(\cc)=-\lambda(\alpha,\beta)\nu\left[f^\tr(\cc)-f_0^\tr(\cc)\right]
+\frac{\zt}{2}\frac{\partial}{\partial \cc}\cdot\left[\VV f^\tr(\cc)\right],
\label{3.1}
\eeq
\beq
\partial_t f^\rot(\ww)+\mathbf{u}\cdot \nabla f^\rot(\ww)=-\lambda(\alpha,\beta)\nu\left[f^\rot(\ww)-f_0^\rot(\ww)\right]
+\frac{1}{2}\frac{\partial}{\partial \ww}\cdot\left\{\left[2\zeta_\Omega\bm{\Omega} +{\overline{\xi}^\rot}\left(\ww-\bm{\Omega}\right)\right]f^\rot(\ww)\right\}.
\label{3.2}
\eeq
Upon writing Eq.\ \eqref{3.2} we have taken into account that $\partial_t n+\nabla\cdot(n\mathbf{u})=0$ and have introduced the \emph{approximation}
\beq
\int d\cc\, \cc f(\cc,\ww)\to n\mathbf{u}f^\rot(\ww).
\label{3.3}
\eeq
Thanks to this approximation, the kinetic model \eqref{2.17} for the joint distribution $f$ yields the set of two coupled kinetic equations \eqref{3.1} and \eqref{3.2}.

Actually, the kinetic equation \eqref{3.1} for the translational distribution function $f^\tr$ is coupled, via $\zt$, to the rotational distribution $f^\rot$  only through the first and second moments $\bm{\Omega}$ and $\Tr$. Therefore, only the equations for these two quantities are needed to close Eq.\ \eqref{3.1}. {}From Eq.\ \eqref{3.2} one gets
\beq
\partial_t \bm{\Omega}+\mathbf{u}\cdot\nabla \bm{\Omega}=-\zeta_\Omega \bm{\Omega},\quad
\partial_t \Tr+\mathbf{u}\cdot\nabla\Tr=-\zr \Tr.
\label{3.4}
\eeq
Equation \eqref{3.4} can alternatively be derived directly from Eq.\ \eqref{2.1} by applying the approximations
\beq
\int d\cc\int d\ww\, \cc \ww f\to n\mathbf{u}\bm{\Omega},\quad \frac{I}{3}\int d\cc\int d\ww\, \cc \omega^2 f\to n\mathbf{u}\Tr,
\label{3.5}
\eeq
which are weaker than Eq.\ \eqref{3.3}.

The simple BGK-like model \eqref{3.1} is formally analogous to the one proposed in Ref.\ \cite{BDS99} for smooth spheres, Eq.\ \eqref{1.14}. The key difference is that the cooling rate $\zeta$ is replaced by the energy production rate $\zt$, which depends not only on $\Tt$ but also on $\bm{\Omega}$ and $\Tr$, the latter two quantities obeying Eq.\ \eqref{3.4}. Despite its crudeness, this model can be useful to explore the basic influence of roughness on the translational properties of a dilute granular gas.

\section{Application to the simple shear flow}

\begin{figure}
  \includegraphics[height=.25\textheight]{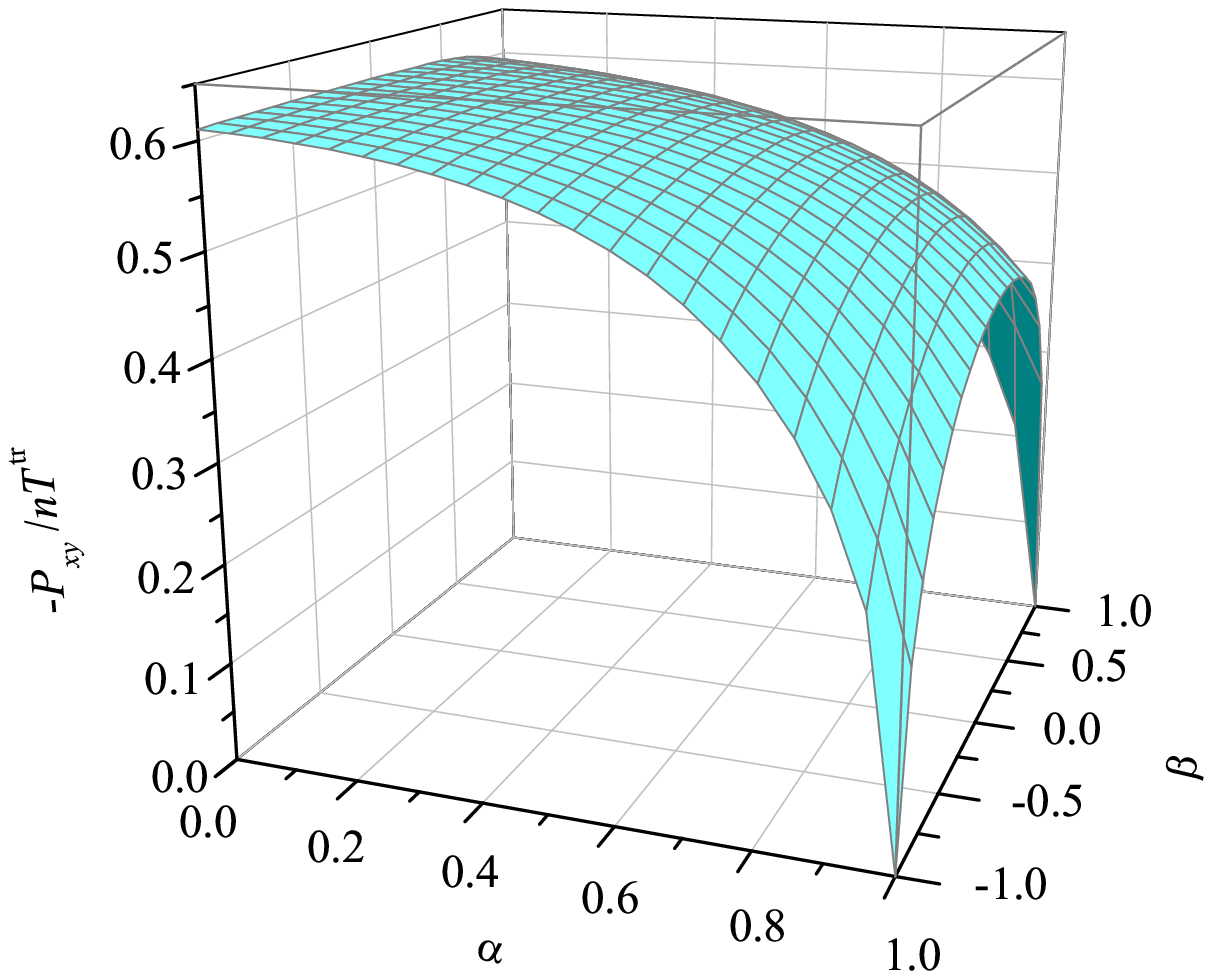} \hspace{2.5cm} \includegraphics[height=.25\textheight]{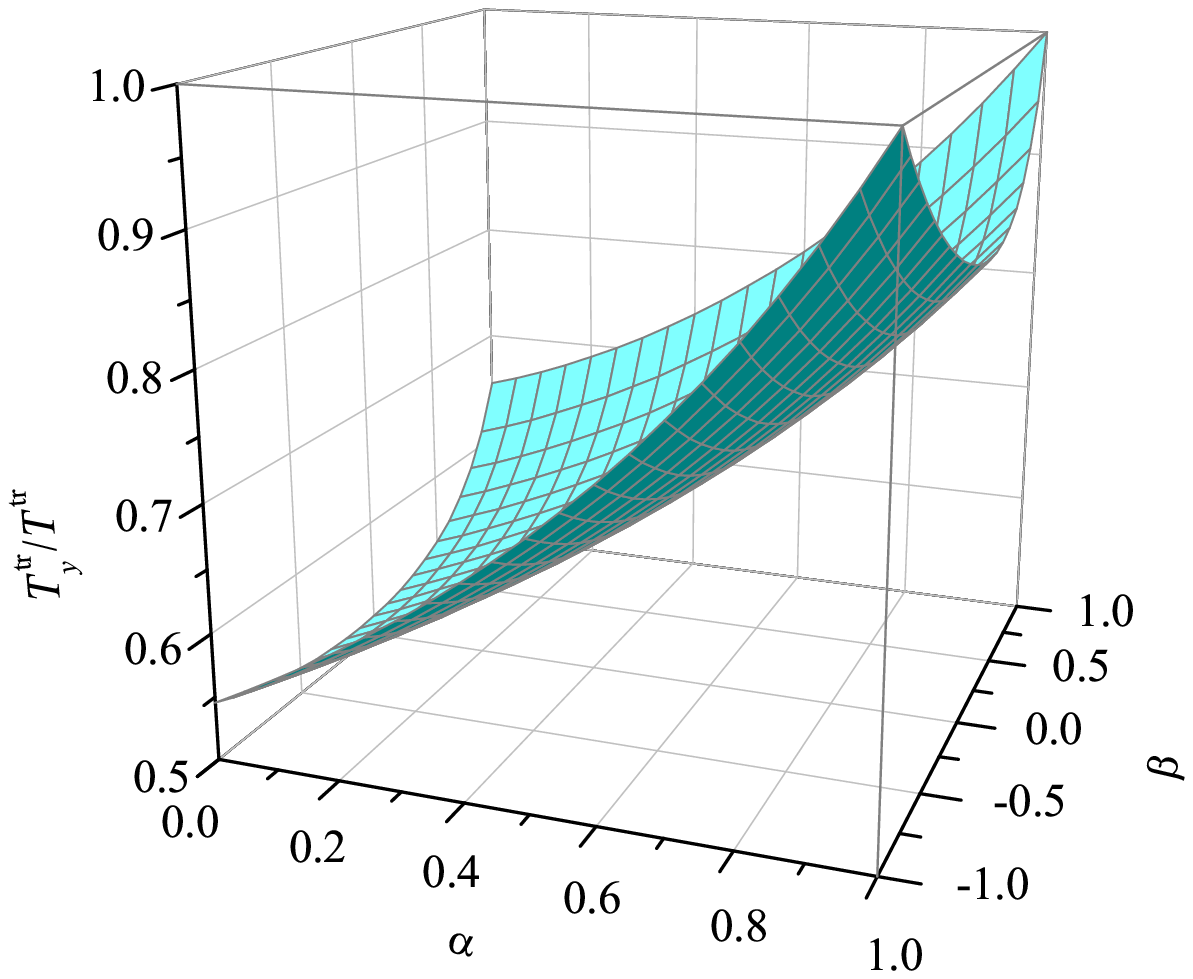}
\caption{Surface plots of $-P_{xy}/n\Tt$ (left) and $\Tt_y/\Tt$ (right) as functions of $\alpha$ and $\beta$ in the simple shear flow of a dilute gas of inelastic rough hard spheres with a uniform mass distribution ($\kappa=\frac{2}{5}$).
\label{fig1}}
\end{figure}

\begin{figure}
  \includegraphics[height=.25\textheight]{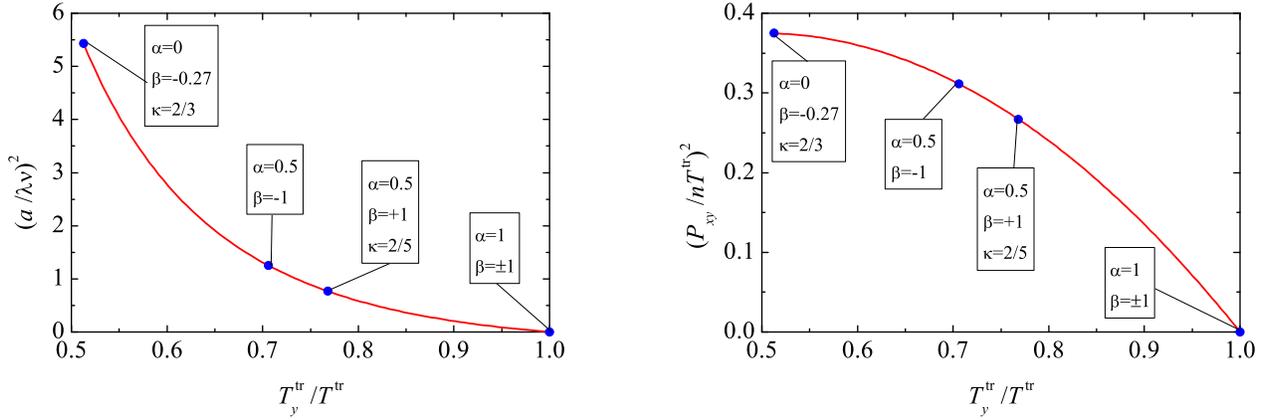}
\caption{Parametric plots of $(a/\lambda \nu)^2$ and $(P_{xy}/n\Tt)^2$ as functions of $\Tt_y/\Tt$ in the simple shear flow of a dilute gas of inelastic rough hard spheres. The circles indicate four representative cases.
\label{fig2}}
\end{figure}

\begin{figure}
  \includegraphics[height=.25\textheight]{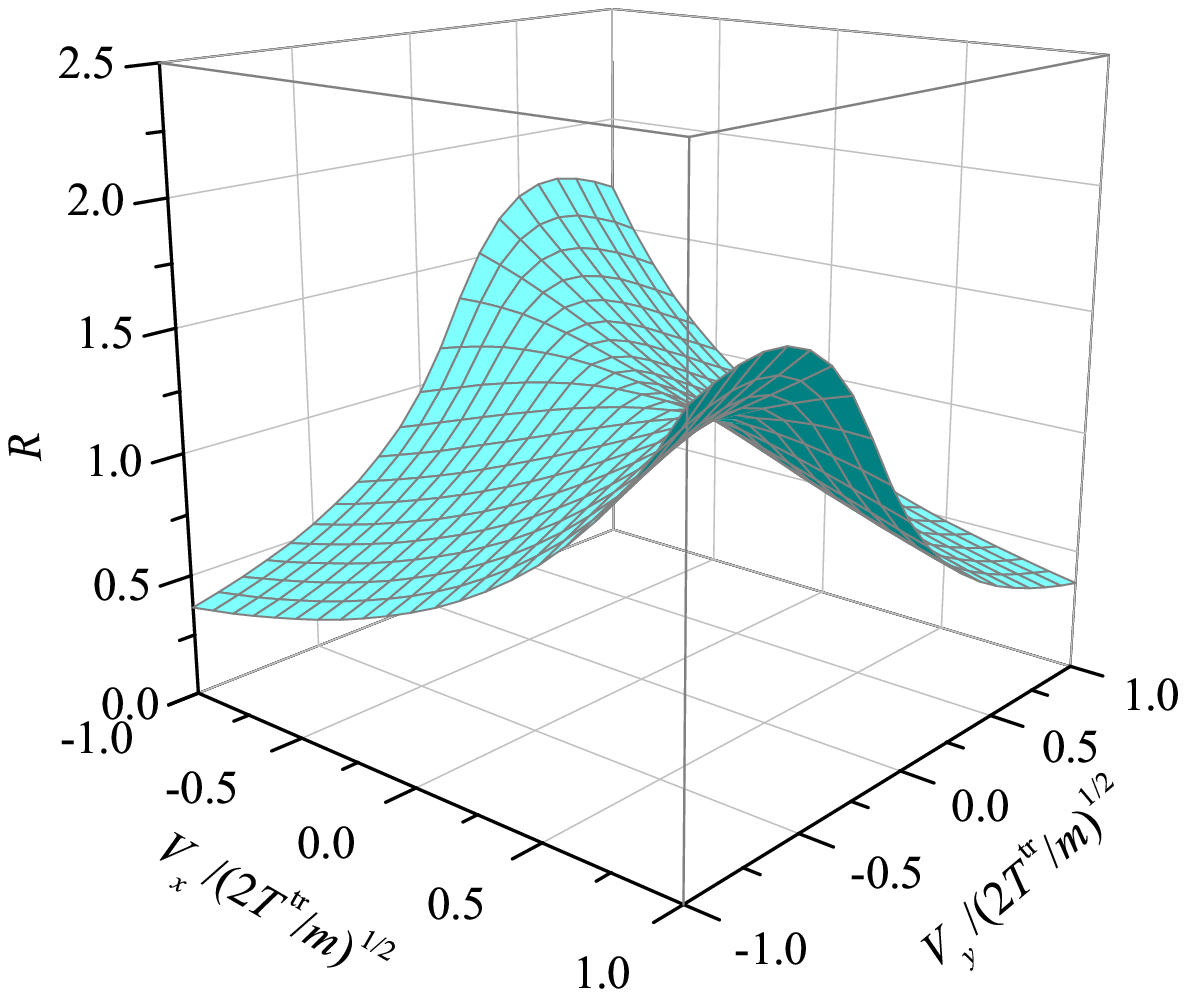} \hspace{2.5cm} \includegraphics[height=.25\textheight]{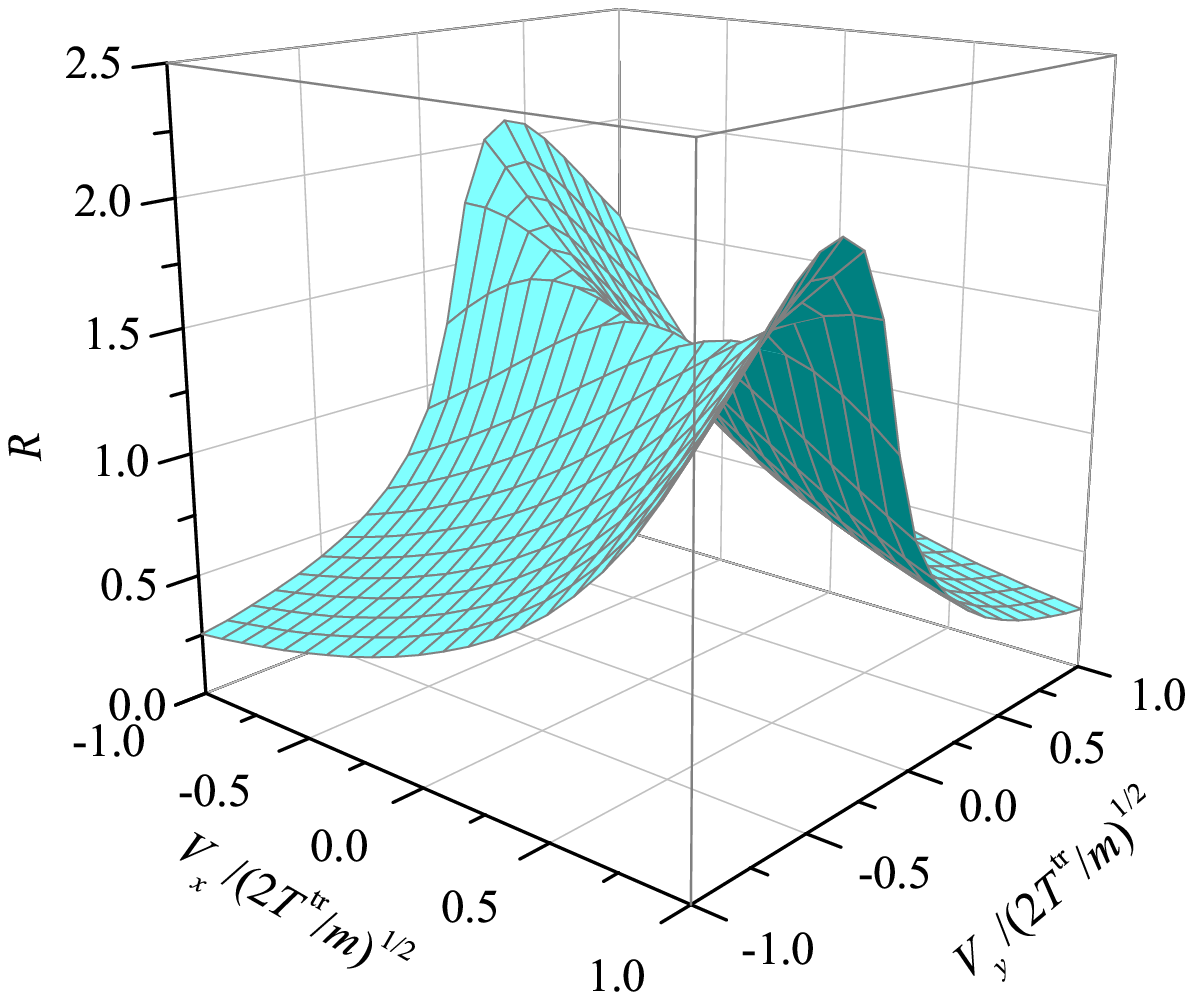}
\caption{Surface plots of $R(V_x,V_y)\equiv {g}^\tr(V_x,V_y)/{g}_0^\tr(V_x,V_y)$ for $\alpha=0.8$ and $\beta=-1$ (left) and for $\alpha=0.8$, $\beta=0.2$, and $\kappa=\frac{2}{5}$ (right) in the simple shear flow of a dilute gas of inelastic rough hard spheres.
\label{fig3}}
\end{figure}

The simple (or uniform) shear flow is an
{incompressible} flow defined by a linear velocity
field $\mathbf{u}=ay\widehat{\mathbf{x}}$, a uniform  density, and a uniform  temperature. This  paradigmatic state is macroscopically quite
simple since only a hydrodynamic gradient exists
($a =\partial u_x/\partial y$) and moreover it is a constant. On the other hand, it is important to remark that the  steady state resulting from the balance
between inelastic cooling and viscous heating is
inherently non-Newtonian \cite{SGD04}. In the steady state Eq.\ \eqref{3.1} becomes \cite{SA05,SGD04}
\beq
-a V_y\frac{\partial}{\partial V_x} f^\tr(\VV)=-\lambda(\alpha,\beta)\nu\left[f^\tr(\VV)-f_0^\tr(\VV)\right]
+\frac{\zt}{2}\frac{\partial}{\partial \VV}\cdot\left[\VV f^\tr(\VV)\right].
\label{4.1}
\eeq
Taking into account that $\partial_t+\mathbf{u}\cdot\nabla=0$ in the steady simple shear flow, Eq.\ \eqref{3.4} yields $\bm{\Omega}=\mathbf{0}$, $\zr=0$. The latter condition determines the ratio between the rotational and translational temperatures from Eq.\ \eqref{2.13}. Insertion of this ratio into Eq.\ \eqref{2.12} gives $\zt$. The results are
\beq
\frac{\Tr}{\Tt}=\q\frac{1+\beta}{1-\beta+2\q}, \quad \zt=\frac{5}{12}\left(1-\alpha^2+2\q\frac{1-\beta^2}{1-\beta+2\q}\right)\nu.
\label{4.2}
\eeq
Interestingly, the temperature ratio is independent of $\alpha$. It ranges from $\Tr/\Tt=0$ in the perfectly smooth case ($\beta=-1$) to  $\Tr/\Tt=1$ in the perfectly rough case ($\beta=1$).
This temperature ratio is the same as in the homogeneous steady state driven by a white-noise thermostat.
Multiplying both sides of Eq.\ \eqref{4.1} by $V_iV_j$ and integrating over velocity one gets a set of algebraic coupled linear equations for the elements $P_{ij}=m\int d\VV\, V_i V_j \ft(\VV)$ of the pressure tensor. The structure is the same as in the smooth case \cite{SA05,BRM97} and so only the final results are quoted here:
\beq
\Tt=\frac{2ma^2}{3\pi\lambda^2(16\sigma^2n/5)^2}\frac{1}{\widetilde{\xi}^\tr\left(1+\widetilde{\xi}^\tr\right)^2},\quad
\frac{P_{xy}}{n\Tt}=-\frac{\sqrt{3\widetilde{\xi}^\tr/2}}{1+\widetilde{\xi}^\tr},\quad
\frac{T_y^\tr}{\Tt}= \frac{T_z^\tr}{\Tt}=\frac{1}{1+\widetilde{\xi}^\tr},\quad
 \frac{T_x^\tr}{\Tt}= 3-2\frac{T_y^\tr}{\Tt},
\label{4.3}
\eeq
where $\widetilde{\xi}^\tr\equiv \zt/\lambda\nu$ is the scaled energy production rate and we have introduced the anisotropic temperatures $\Tt_i\equiv P_{ii}/n$, where $P_{ii}$ is the $i$th normal stress.
The absolute maximum value of $\widetilde{\xi}^\tr$ is $\frac{5}{294}\left(109-16\sqrt{11}\right)\simeq 0.95$ and corresponds to $\alpha=0$, $\kappa=\frac{2}{3}$ and $\beta=13-4\sqrt{11}\simeq -0.27$.
Elimination of $\widetilde{\xi}^\tr$ between $\Tt_y/\Tt$ and either $a^2/\Tt$ or $P_{xy}/n\Tt$  yields two nonequilibrium ``equations of state'' independent of $\alpha$ and $\beta$, namely
\beq
\widetilde{a}^2\equiv \left(\frac{a}{\lambda \nu}\right)^2=\frac{3}{2}\left(\frac{\Tt}{\Tt_y}\right)^2\left(\frac{\Tt}{\Tt_y}-1\right)
,\quad
\left(\frac{P_{xy}}{n\Tt}\right)^2=\frac{3}{2}\frac{\Tt_y}{\Tt}\left(1-\frac{\Tt_y}{\Tt}\right).
\label{4.4}
\eeq

Figure \ref{fig1} shows the dependence of the ratios $-P_{xy}/n\Tt$ and $\Tt_y/\Tt$ on both coefficients of restitution $\alpha$ and $\beta$ for the case $\q=\frac{2}{5}$ (uniform spherical mass distribution). The anisotropic effects induced by the shearing are measured by the departure of $P_{xy}/n\Tt$ and $\Tt_y/\Tt$ from $0$ and $1$, respectively. We can observe that, for a fixed value of $\beta$, those effects increase monotonically with decreasing $\alpha$. On the other hand, for a fixed value of $\alpha$ the influence of roughness is not monotonic, the higher deviations from anisotropy taking place for central values of $\beta$. The dependence on $\beta$ becomes less important as the inelasticity increases. In the elastic case ($\alpha=1$) and $\beta=\pm 1$, $P_{xy}/n\Tt=0$ and $\Tt_y/\Tt=1$, as expected.
The two universal equations of state are plotted in Fig.\ \ref{fig2}. Every triad $(\alpha,\beta,\kappa)$ is represented by one single point in each graph and all the representative points lie on the curves given by Eq.\ \eqref{4.4}. The end points of the curves correspond to $(\alpha=0,\beta\simeq -0.27,\kappa=\frac{2}{3})$ and $(\alpha=1,\beta=\pm 1,\kappa=\text{arbitrary})$.

A practical advantage of kinetic models is the possibility
of obtaining explicitly the velocity distribution function. Exploiting the analogy with the smooth case \cite{SA05}, the
solution to Eq.\ \eqref{4.1} turns out to be
 \beq
\ft(\VV)=\int ds\, e^{-\left(1-\frac{3}{2}\widetilde{\xi}^\tr\right)s}\ft_0\left(e^{\frac{1}{2}\widetilde{\xi}^\tr s}
\left(\VV+\widetilde{a}sV_y\widehat{\mathbf{x}}\right)\right).
\label{4.5}
\eeq
This expression is formally analogous to that of an ordinary fluid with a Gaussian thermostat \cite{GS03}.
The marginal distribution ${g}^\tr(V_x,V_y)\equiv\int_{-\infty}^\infty dV_z\, \ft(\VV)$ is also given by Eq.\ \eqref{4.5} except for the replacements $\frac{3}{2}\widetilde{\xi}^\tr\to \widetilde{\xi}^\tr$, $\ft_0(\VV)\to {g}_0^\tr(V_x,V_y)\equiv\int_{-\infty}^\infty dV_z\, \ft_0(\VV)$. Figure \ref{fig3} show the ratio $R(V_x,V_y)\equiv{g}^\tr(V_x,V_y)/{g}_0^\tr(V_x,V_y)$ for inelastic smooth spheres with $\alpha=0.8$ and for inelastic rough spheres with $\alpha=0.8$, $\beta=0.2$, and $\kappa=\frac{2}{5}$. As can be observed, the distortion from the local equilibrium distribution is higher in the latter case than in the former.

\section{Concluding remarks}
In this paper a two-level BGK-like description for a dilute granular gas of inelastic rough hard spheres has been proposed. At a more fundamental level, the  one-body joint distribution function $f(\cc,\ww)$ is assumed to obey the kinetic equation \eqref{2.1} with the replacement \eqref{2.17}. The model preserves the Boltzmann collisional integrals \eqref{2.5}--\eqref{2.6} with the de-spinning rate ($\zeta_\Omega$) and the partial energy production rates ($\zt$ and $\zr$) given by Eqs.\ \eqref{2.15}--\eqref{2.13}. At a simpler and cruder level, the translational distribution function $\ft(\cc)$ is assumed to obey Eq.\ \eqref{3.1}, complemented by Eq.\ \eqref{3.4}, while the rotational distribution $\fr(\ww)$ satisfies Eq.\ \eqref{3.2}. The second level prevents one from accounting for correlations between the translational and rotational degrees of freedom \cite{KBPZ09}. On the other hand, the model kinetic equation \eqref{3.1} can be useful to investigate the basic influence of roughness on  the translational properties of the granular gas.

The kinetic model made of Eqs.\ \eqref{3.1} and \eqref{3.4} has been applied to the steady simple shear flow problem. The solution predicts that the temperature ratio $\Tr/\Tt$ is independent of the coefficient of normal restitution $\alpha$ and is given by Eq.\ \eqref{4.2}. On the other hand, according to Eq.\ \eqref{4.3}, the reduced shear stress $P_{xy}/n\Tt$, the reduced shear rate $a/\lambda\nu$, and the anisotropic temperature ratios $\Tt_x/\Tt$ and $\Tt_y/\Tt$ depend on $\alpha$ and $\beta$ only through the reduced energy production rate $\widetilde{\xi}^\tr$. It is observed that, at a given value of $\alpha$, the dependence of those quantities on $\beta$ is not monotonic, this effect being less pronounced  as inelasticity increases.

It is planned to carry out computer simulations to test the above theoretical predictions for the simple shear flow. Moreover, the kinetic model \eqref{3.1} will be applied to other states that have been solved in the case of perfectly smooth spheres, such as the Couette flow \cite{TTMGSD01}, the gravity-driven Poiseuille flow \cite{TS04}, and the uniform longitudinal flow \cite{S08}.

%%%%%%%%%%%%%%%%%%%%%%%%%%%%%%%%%%%%%%%%%%%%%%%%
%% BACKMATTER
%%%%%%%%%%%%%%%%%%%%%%%%%%%%%%%%%%%%%%%%%%%%%%%%

\begin{theacknowledgments}
The author is grateful to V. Garz\'o and G. M. Kremer for insightful discussions.   This work has been supported by the
Ministerio de Educaci\'on y Ciencia (Spain) through Grant No.\
FIS2007--60977 (partially financed by FEDER funds) and by the Junta
de Extremadura (Spain) through Grant No.\ GR10158.
\end{theacknowledgments}



\bibliographystyle{aipproc}   % if natbib is available

\end{document}